\documentclass[10pt,conference]{IEEEtran} 

\usepackage{color}
\usepackage{balance}
\usepackage{xcolor}
\usepackage{multirow}

\usepackage{dblfloatfix}
\usepackage{booktabs} 
\usepackage[export]{adjustbox}
\usepackage{url}
\usepackage{cite}
\usepackage{hhline}
\usepackage{amssymb}
\usepackage{amsfonts}
\usepackage{amsmath}
\usepackage{hyperref}
\usepackage{algorithm}
\usepackage{algorithmic}
\usepackage{pbox}

\newcommand{\overbar}[1]{\mkern 1.5mu\overline{\mkern-1.5mu#1\mkern-1.5mu}\mkern 1.5mu}

\definecolor{Gray}{gray}{0.95}
\definecolor{LightGray}{gray}{0.975}

\newcommand{\bi}{\begin{itemize}[leftmargin=0.4cm]}
	\newcommand{\ei}{\end{itemize}}
\newcommand{\be}{\begin{enumerate}}
	\newcommand{\ee}{\end{enumerate}}

\newcommand{\uniquePR}{190 }
\newcommand{\newPR}{170 }
\newcommand{\totalPR}{210 }

\colorlet{shadecolor}{blue!20}

\definecolor{Gray}{rgb}{0.88,1,1}
\definecolor{Gray}{gray}{0.85}
\makeatletter
\let\th@plain\relax
\makeatother
\usepackage[framed]{ntheorem}
\usepackage{framed}
\usepackage{tikz}
\usetikzlibrary{shadows}

\theoremclass{insight}
\theoremstyle{break}
\tikzstyle{thmbox} = [rectangle, rounded corners, draw=black,
 fill=Gray,  drop shadow={fill=black, opacity=1}]

\newshadedtheorem{insight}{Insight}

\begin{document}
\title{ Replication Can Improve Prior   Results:\\ A GitHub  Study of Pull Request Acceptance} 
\author{Di Chen, Kathyrn Stolee,  Tim Menzies\\
Computer Science, NC State, USA,\\
dchen20@ncsu.edu;  ktstolee@ncsu.edu;  timm@ieee.org}
\maketitle


\begin{abstract}


Crowdsourcing and data mining can be used to effectively reduce the effort associated with the partial replication and enhancement of qualitative studies.
 
For example, in a primary study, other researchers explored factors influencing the fate of GitHub pull requests using an extensive qualitative analysis of 20 pull requests. Guided by their findings, we mapped some of their qualitative insights onto quantitative questions. To determine how well their findings generalize, we collected much more data (170 additional pull requests from 142 GitHub projects). Using crowdsourcing, that data was augmented with subjective qualitative human opinions about how pull requests extended the original issue. The crowd's answers were then combined with quantitative features and, using data mining, used to build a   predictor for whether code would be merged. That predictor was far more accurate that one built from the primary study's qualitative factors (F1=90 vs 68\%),  illustrating the value of a mixed-methods approach and replication to improve prior results.

To test the generality of this approach, the next step in future work is to conduct other studies that extend qualitative studies with crowdsourcing and data mining.


\end{abstract}

\pagestyle{plain}
\section{Introduction}

Our ability to generate models from software engineering   data has out-paced our abilities
to reflect on those models.
Studies can use 
thousands of projects, 
millions of lines of code, or
tens of thousands of programmers~\cite{ray2014large}.
However, when insights from human experts
are overlooked, the conclusions from the
automatically generated models can be both wrong and
misleading~\cite{o2016weapons}.  
After observing 
case studies where data mining in software engineering
led to spectacularly wrong results, Basili and Shull~\cite{shull02} 
recommend qualitative analysis to collect and use insights from subject
matter experts who understand software engineering.

The general problem we explore is how partial replication studies can scale and deepen the insights gained from primary qualitative studies. 
That is, after collecting qualitative insights from an in-depth analysis of a small sample, a partial replication study is conducted using a subset of the insights as a guide, but targeting a larger sample and using a different empirical methodology. 
To show that a given result is robust, the ideal case is for a completely independent set of researchers to replicate a published study using their own experimental design~\cite{Shull:2008:RRE:1361580.1361587}. 
In this work, we explore a mixed-methods approach using a crowdsourced evaluation and  data mining to build on a primary qualitative study from prior work~\cite{tsay2014let}, aimed at the goal of understanding the factors that govern pull request acceptance. 

Crowdsourcing brings advantages of a lower cost compared to professional experts. Micro-task crowdsourcing using established platforms such as Amazon's Mechanical Turk (MTurk)~\cite{mturk} also provides a large worker pool with great diversity and fast completion times. The main issue with crowdsourcing is the low quality; an uncontrolled experimental context often leads to less credible results; however, some results suggest that crowdsourced workers perform similarly to student populations~\cite{stoleeesem2013}. 
Data mining, on the other hand, is good at predicting future patterns based on the past. It is an inexpensive and fast tool to analyze quantitative data from crowdsourcing results, especially when data is large. 
However, data mining is limited in that it looks narrowly at the data.
The starting point of this investigation was a conjecture that combining
crowdsourcing and data mining would lead to better results that using either separately.

To test this conjecture, we used
Tsay,   Dabbish,     Herbsleb (hereafter, TDH)~\cite{tsay2014let}. 
That study
 explored how GitHub-based teams debate what new code gets merged through the lens of pull requests.
To do this, they used a 
labor-intensive qualitative interview-process of 47 users of GitHub, 
as well as in-depth case studies of 20 pull-requests. They found that the submitter's level of prior interaction on the project changed how core and audience members interacted with the submitter during discussions around contributions. The results provide many insights into the factors and features that govern pull request acceptance. 
The TDH authors were clear about their methodology and results, making it a good candidate for partial replication and extension. 

  This paper extends  that  primary qualitative study of pull requests with an independent partial replication study using a crowdsourced evaluation and data mining. 
To perform this independent, partial replication and extension, using the insights from the original study, we design questions that can be answered by a a crowdsourced worker pool and serve to confirm some of the original findings. The crowd is able to handle a larger pool of artifacts than the original study, which tests the external validity of the findings. In addition to the original 20 pull requests, the crowd in our study analyzes and additional \newPR pull requests. 
Next, data mining was applied to the crowd's responses, resulting in accurate predictors for pull request acceptance. 
  The predictors based on crowd data were compared to predictors
built using quantitative methods from the literature (i.e., traditional data mining without crowdsourcing and without insights from the primary study).

Our results show that the crowd can replicate the TDH results, and that for those factors studied, most results are stable when scaled to larger data sets. 
After using data mining to develop predictors for pull request acceptance, we found that the predictors based on quantitative factors from the literature were more accurate than the predictor based on the TDH features we studied.
Even though the predictors for pull request acceptance based on data mining were more accurate than predictors based on crowd-generated data alone, it would be extremely premature to use this one result to make generalizations about the relative merits of the different empirical approaches. 
The primary study revealed insights that we were not able to scale up; for example, the original study found that evaluation outcomes of pull request were sometimes more complex than acceptance or rejection. That is, a rejected pull request may be followed up with another pull request from the core team fulfilling the goals of the rejected pull request~\cite{tsay2014let}. 
Such insights would be difficult, if not impossible, to exposed through data mining alone.
That said, other insights can be verified through scaled replication, such as the impact of different features of a pull request discussion on that pull request's acceptance, which we explore in this work.

This paper makes three specific contributions:
\begin{itemize}
    \item A cost-effective, independent, partial replication and extension of a primary study of pull request acceptance factors using a scaled sample of artifacts (RQ1).
    \item Analysis of the external validity of findings from the original study, demonstrating stability in most of the results. This has implications for which questions warrant further analysis (RQ2). 
    \item Comparison of qualitative and quantitative factors that impact pull request acceptance from related work (RQ3). 
\end{itemize}

    To assist other researchers, a reproduction package with all
    our scripts and data is available\footnote{ https://github.com/dichen001/IST\_17}
     and in archival
    form (with a DOI)\footnote{https://doi.org/10.5281/zenodo.802698}
     to simplify all future citations to this material.


\section{Background} \label{background}

    \begin{figure}[!t]
			\centering
			\includegraphics[width=\columnwidth]{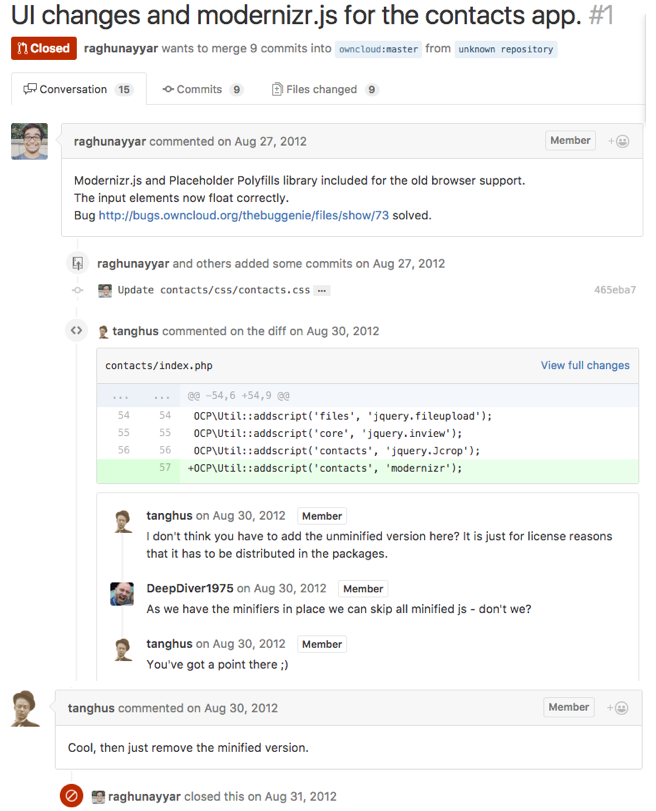}
			\vspace{-12pt}
			\caption{An example GitHub pull request}
			\vspace{-12pt}
			\label{fig:pr}
		\end{figure}

    With over 14 million users and 35 million repositories as of April 2016, GitHub has become the largest and most influential open source projects hosting sites.
    Numerous qualitative~\cite{ dabbish2012social, tsay2014let, gousios2015work, gousios2016work, marlow2013impression, begel2013social,mcdonald2013performance, pham2013creating},   
    quantitative~\cite{tsay2014influence, gousios2013exploration, tsay2012social, yu2015wait, zhang2014investigating, rahman2014insight, takhteyev2010investigating, thung2013network, ray2014large} 
    and mixed methods studies~\cite{kalliamvakou2014promises, blincoe2016understanding} 
    have been published about GitHub. 

    Pull requests are created when 
    contributors want their changes to be merged to the main repository. 
    After core members get pull requests, 
    they inspect the changes and 
    decide whether to accept or reject them. 
    This process usually involves code inspection, 
    discussion, and inline comments between contributors and repository owners. 
    Core members who have the ability to close the pull requests 
    by either accepting the code and merging the contribution with the master branch, 
    or rejecting the pull requests. 
    Core members could also ignore the pull requests and leave them in an open state.

     \begin{table*}[!b]
\centering
\caption{A sample of related qualitative and quantitative work.
Here, by ``quantitative'', we mean using data mining with little
to no interaction with project personnel.}
  \label{tab:lit_rev}
\begin{adjustbox}{max width=\linewidth}
\begin{tabular}{|l|l|l|p{2in}|p{3in}|}
\hline
\textbf{Year}                     & \textbf{Source} & \textbf{Method}              & \textbf{Data}                                                          & \textbf{Title}                                                                                  \\ \hline
2012~\cite{dabbish2012social}        & CSCW            & Qualitative                  & Interview 24 GitHub Users.  Pull requests case study 10.               & Social coding in GitHub: transparency and collaboration in an open software repository          \\ \hline
2013~\cite{marlow2013impression}     & CSCW            & Qualitative                  & Interview 18 GitHub users.  Pull requests case study 10.               & Impression Formation in Online Peer Production: Activity Traces and Personal Profiles in GitHub \\ \hline
2014~\cite{tsay2014let}              & FSE             & Qualitative                  & Interview 47 GitHub users.  Pull requests case study 20.               & \textbf{Let's Talk About It: Evaluating Contributions through Discussion in GitHub (TDH)}                      \\ \hline
2015~\cite{gousios2015work}          & ICSE            & Qualitative                  & Online survey 749 integrators.                                         & Work Practices and Challenges in Pull-Based Development: The Integrator’s Perspective           \\ \hline
2016~\cite{gousios2016work}          & ICSE            & Qualitative                  & Online survey 645 contributors.                                        & Work Practices and Challenges in Pull-Based Development: The Contributor’s Perspective          \\ \hline
2014~\cite{gousios2014exploratory}   & ICSE            & Quantitative                 & GHTorrent, 166,884 pull requests                                       & An Exploratory Study of the Pull-Based Software Development Model                               \\ \hline
2014~\cite{tsay2014influence}        & ICSE            & Quantitative                 & GitHub API, GitHub Archive. 659,501 pull requests                      & Influence of Social and Technical Factors for Evaluating Contribution in GitHub.                \\ \hline
2014~\cite{yu2014reviewer}           & ICSME           & Quantitative                 & GHTorrent, 1,000 pull requests.                                        & Reviewer Recommender of Pull-Requests in GitHub                                                 \\ \hline
2014~\cite{vasilescu2014continuous}  & ICSME           & Quantitative                 & GHTorrent                                                              & Continuous Integration in a SocialCoding World Empirical Evidence from GitHub                   \\ \hline
2014~\cite{yu2014should}             & APSEC           & Quantitative                 & GHTorrent, 1,000 pull requests.                                        & Who Should Review This Pull-Request: Reviewer Recommendation to Expedite Crowd Collaboration    \\ \hline
2014~\cite{zhang2014investigating}   & CrowdSoft       & Quantitative                 & GHTorrent, GitHubArchive.                                              & Investigating Social Media in GitHub’s Pull-Requests: A Case Study on Ruby on Rails             \\ \hline
2014~\cite{gousios2014dataset}       & MSR             & Quantitative                 & GHTorrent                                                              & A Dataset for Pull-Based Development Research                                                   \\ \hline
2014~\cite{rahman2014insight}        & MSR             & Quantitative                 & GHTorrent,  78,955 pull requests.                                      & An Insight into the Pull Requests of GitHub                                                     \\ \hline
2014~\cite{pletea2014security}       & MSR             & Quantitative                 & GHTorrent,  54,892 pull requests.                                      & Security and emotion sentiment analysis of security discussions on GitHub                       \\ \hline
2014~\cite{brunet2014developers}     & MSR             & Quantitative                 & GHTorrent                                                              & Do developers discuss design                                                                    \\ \hline
2014~\cite{padhye2014study}          & MSR             & Quantitative                 & GHTorrent, 75,526 pull requests.                                       & A study of external community contribution to opensource projects on GitHub                     \\ \hline
2015~\cite{van2015automatically}     & MSR             & Quantitative                 & GHTorrent                                                              & Automatically Prioritizing Pull Requests                                                        \\ \hline
2015~\cite{yu2015wait}               & MSR             & Quantitative                 & GHTorrent, 103,284 pull requests.                                      & Wait For It: Determinants of Pull Request Evaluation Latency on GitHub                          \\ \hline
2014~\cite{kalliamvakou2014promises} & MSR             & \pbox{20cm}{ Quantitative \& \\ Qualitative}  & \pbox{20cm}{ Quant. : GHTorrent \\ Qual. : 240 Survey, 434 projects.} & The promises and perils of mining GitHub                                                        \\ \hline
\end{tabular}
\end{adjustbox}

\end{table*}

    Figure~\ref{fig:pr} shows an example of pull requests with reduced discussion (No.16 of the 20 pull requests TDH studied.\footnote{\url{http://www.jsntsay.com/work/FSE2014.html}}), inline code comments and final result (closed).  
    This pull request, on the topic of {\em UI changes and moderizr.js for the contacts app}, was submitted by user {\em raghunayyar} on August 27, 2012. The intention of the pull request was to address a bug, which is linked within the submitter's first comment. User {\em tanghus} commented on the diff for file {\tt contacts/index.php} on August 30, 2012. A discussion ensued between user {\em tanghus} and user {\em DeepDiver1975}. After one more comment from {\em tanghus}, the pull request is closed by the submitter on August 31, 2012. 
    All three people involved with the pull request are {\em Member}s, meaning. From the contributors page of this repository, we and the crowd could find the user {\em  tanghus} and user {\em DeepDiver1975} are core members, while the submitter, user {\em raghunayyar}, is an external developer.
    The pull request was closed without being merged. 
  The crowd worker who was assigned to analyze this pull requests finds: 
  \begin{enumerate}
  \item  There are core developers supporting this pull requests. 
  \item
  There are alternates proposed by the core members. 
  \item
  There are people disapproving the proposed solution . 
  \item No one disapproves the problem being solved here. 
  \item The pull requests are rejected but the core team implemented their own solution to the problem in the contribution.
  \end{enumerate}
  Note that  findings 3 and 4 conflict with the results from TDH. TDH finds there are disapproving comments for the problems being proposed due to project appropriateness
  (but not for the solutions itself being inconsistent, as found by the crowd). 
A small number of such inconsistencies are not unexpected in qualitative work, and replication can help identify where ambiguities may be present.

    In this paper, crowd workers analyze pull request features and the conversations within the pull request, as just illustrated, and answer quantitative questions about the pull request conversations and outcomes.

 \section{Research Questions}
    We evaluate the following research questions: \\
    
      \noindent {\bf  RQ1:} {\em How can we use crowdsourcing to partially replicate TDH's work with high quality?}
      One challenge with crowdsourcing is quality control; we employ several strategies to encourage high-quality responses from the crowd to determine if the crowd can partially replicate the results from the TDH paper. 
      \\
       
	  \noindent {\bf  RQ2:} {\em Does crowdsourcing identify which conclusions from the primary study are stable?} 
	This question is important for the external validity of the original findings used in the extension part of the experiment. We collected  \newPR
	  additional pull requests using similar sampling criteria to the primary study, and added these to the original 20. We then tested if the crowd reaches the same or different conclusions using the original 20 and using the extended data set.  
	  \\
	  
	  \noindent {\bf  RQ3:} {\em Can the pull request features identified in the primary study accurately predict pull request acceptance? } 
	  Given the larger data set collected and evaluated in this work, there is now an opportunity to evaluate the performance of prediction models based on (1) the subset of features identified as important in the primary study and included in our partial replication (collected in RQ1 and RQ2), and (2) features identified as important in previous data mining-only studies (identified from related work). 

            \begin{table*}[!b]
\centering
\caption{Features Used in Related Work. $\square$ indicates that a feature is used;  $\blacksquare$ indicated the feature is found to be heavily related to the results of pull requests in the corresponding paper.}

  \label{all_features}
\begin{adjustbox}{max width=\linewidth}
\begin{tabular}{llp{3.4in}ccccccc}
\textbf{Category}      & \textbf{Fetures}   & \textbf{Description}                                                                & \textbf{~\cite{gousios2014dataset}} & \textbf{~\cite{gousios2014exploratory}} & \textbf{~\cite{tsay2014influence}} & \textbf{~\cite{yu2015wait}} & \textbf{~\cite{zhang2014investigating}} & \textbf{~\cite{rahman2014insight}} & \textbf{Ours}  \\ \hline
\multicolumn{1}{l|}{Pull Request} & lifetime\_minites             & Minutes between opening and closing                                                            & $\square$                                  &                                                &                                           &                                    &                                                &                                           &                \\
\multicolumn{1}{l|}{Pull Request} & mergetime\_minutes            & Minutes between opening and merging                            & $\square$                                  &                                                &                                           &                                    &                                                &                                           &                \\
\multicolumn{1}{l|}{Pull Request} & num\_commits                  & Number of commits                                                                              & $\square$                                  & $\square$                                      & $\square$                                 & $\blacksquare$                     &                                                &                                           & $\square$      \\
\multicolumn{1}{l|}{Pull Request} & src\_churn                    & Number of lines changed (added + deleted)                                                      & $\square$                                  & $\blacksquare$                                 &                                           & $\blacksquare$                     &                                                &                                           & $\square$      \\
\multicolumn{1}{l|}{Pull Request} & test\_churn                   & Number of test lines changed                                                                   & $\square$                                  & $\square$                                      &                                           &                                    &                                                &                                           &                \\
\multicolumn{1}{l|}{Pull Request} & files\_added                  & Number of files added                                                                          & $\square$                                  &                                                &                                           &                                    &                                                &                                           &                \\
\multicolumn{1}{l|}{Pull Request} & files\_deleted                & Number of files deleted                                                                        & $\square$                                  &                                                &                                           &                                    &                                                &                                           &                \\
\multicolumn{1}{l|}{Pull Request} & files\_modified               & Number of files modified                                                                       & $\square$                                  &                                                &                                           &                                    &                                                &                                           &                \\
\multicolumn{1}{l|}{Pull Request} & files\_changed                & Number of files touched (sum of the above)                                                     & $\square$                                  & $\square$                                      & $\square$                                 &                                    &                                                &                                           &                \\
\multicolumn{1}{l|}{Pull Request} & src\_files                    & Number of source code files touched by the pull request                                        & $\square$                                  &                                                &                                           &                                    &                                                &                                           &                \\
\multicolumn{1}{l|}{Pull Request} & doc\_files                    & Number of documentation (markup) files touched                                                 & $\square$                                  &                                                &                                           &                                    &                                                &                                           &                \\
\multicolumn{1}{l|}{Pull Request} & other\_files                  & Number of non-source, non-documentation files touched                                          & $\square$                                  &                                                &                                           &                                    &                                                &                                           &                \\
\multicolumn{1}{l|}{Pull Request} & num\_commit\_comments         & Total number of code review comments                                                       & $\square$                                  &                                                &                                           &                                    &                                                &                                           &                \\
\multicolumn{1}{l|}{Pull Request} & num\_issue\_comments          & Total number of discussion comments                                                        & $\square$                                  &                                                &                                           &                                    &                                                &                                           &                \\
\multicolumn{1}{l|}{Pull Request} & num\_comments                 & Total number of discussion and code review                                      & $\square$                                  & $\square$                                      & $\blacksquare$                            & $\blacksquare$                     &                                                &                                           & $\square$      \\
\multicolumn{1}{l|}{Pull Request} & num\_participants             & Number of participants in the discussion                                                       & $\square$                                  & $\square$                                      &                                           &                                    &                                                &                                           &                \\
\multicolumn{1}{l|}{Pull Request} & test\_inclusion               & Whether or not the pull request included test cases                                            &                                            &                                                & $\square$                                 & $\square$                          &                                                &                                           &                \\
\multicolumn{1}{l|}{Pull Request} & prior\_interaction            & Number of events the submitter has participated previously   &                                            &                                                & $\square$                                 &                                    &                                                &                                           &                \\
\multicolumn{1}{l|}{Pull Request} & social\_distance              & Whether the submitter follows the user who closes the PR                      &                                            &                                                & $\blacksquare$                            & $\square$                          &                                                &                                           & $\square$      \\

\multicolumn{1}{l|}{Pull Request} & strength of social connection & Fraction of members interacted with the submitter in $T_0$       &   &     &   & $\square$         &     &     &      \\
\multicolumn{1}{l|}{ }    &                                & (the last 3 months prior to creation) &&&&&&&\\
\multicolumn{1}{l|}{Pull Request} & description complexity        & Total number of words in the pull request title and description                                &                                            &                                                &                                           & $\square$                          &                                                &                                           &                \\
\multicolumn{1}{l|}{Pull Request} & first human response          & Interval from PR creation to first response by reviewers             &                                            &                                                &                                           & $\blacksquare$                     &                                                &                                           & $\square$      \\
\multicolumn{1}{l|}{Pull Request} & total CI latency:             & Interval from PR creation to the last commit tested by CI            &                                            &                                                &                                           & $\blacksquare$                     &                                                &                                           & $\square$      \\
\multicolumn{1}{l|}{Pull Request} & CI result:                    & Presence of errors and test failures while running Travis-CI                                   &                                            &                                                &                                           & $\blacksquare$                     &                                                &                                           & $\square$      \\
\multicolumn{1}{l|}{Pull Request} & mention-@                     & Weather there exist an @-mention in the comments                                               &                                            &                                                &                                           &                                    & $\square$                                      &                                           &                \\ \hline
\multicolumn{1}{l|}{Repository}   & sloc                          & Executable lines of code at creation time.                                                     & $\square$                                  & $\blacksquare$                                 &                                           &                                    &                                                &                                           & $\square$      \\
\multicolumn{1}{l|}{Repository}   & team\_size                    & Number of active cores in $T_0$                  & $\square$                                  & $\blacksquare$                                 & $\square$                                 & $\square$                          &                                                & $\blacksquare$                            & $\square$      \\
\multicolumn{1}{l|}{Repository}   & perc\_external\_contribs      & Ratio of commits from externals over cores in $T_0$              & $\square$                                  & $\blacksquare$                                 &                                           &                                    &                                                &                                           & $\square$      \\
\multicolumn{1}{l|}{Repository}   & commits\_on\_files\_touched   & Number of total commits on files touched by the PR in $T_0$             & $\square$                                  & $\blacksquare$                                 &                                           & $\square$                          &                                                &                                           & $\blacksquare$ \\
\multicolumn{1}{l|}{Repository}   & test\_lines\_per\_kloc        & Executable lines of test code per 1,000 lines of source code                                   & $\square$                                  & $\blacksquare$                                 &                                           &                                    &                                                &                                           &    $\square$            \\
\multicolumn{1}{l|}{Repository}   & test\_cases\_per\_kloc        & Number of test cases per 1,000 lines of source code                                            & $\square$                                  &                                                &                                           &                                    &                                                &                                           &                \\
\multicolumn{1}{l|}{Repository}   & asserts\_per\_kloc            & Number of assert statements per 1,000 lines of source code                                     & $\square$                                  &                                                &                                           &                                    &                                                &                                           &                \\
\multicolumn{1}{l|}{Repository}   & watchers                      & Project watchers (stars) at creation                                                           & $\square$                                  &                                                & $\blacksquare$                            &                                    &                                                &                                           & $\square$      \\
\multicolumn{1}{l|}{Repository}   & repo\_age                     & Life of a project since the time of data collection                     &                                            &                                                & $\square$                                   & $\square$                          &                                                &                                           &                \\
\multicolumn{1}{l|}{Repository}   & workload                      & Total number of PRs still open at current PR creation time &                                            &                                                &                                           & $\square$                          &                                                &                                           &                \\
\multicolumn{1}{l|}{Repository}   & integrator availability       & Minimal hours until either of the top 2 integrators are active   &                                            &                                                &                                           & $\square$                          &                                                &                                           &                \\
\multicolumn{1}{l|}{Repository}   & project maturity              & Number of forked projects as an estimate of project maturity                 &                                            &                                                &                                           &                                    &                                                & $\blacksquare$                            & $\square$      \\ \hline
\multicolumn{1}{l|}{Developer}    & prev\_pullreqs                & Number of PRs previously submitted by the submitter           & $\square$                                  & $\blacksquare$                                 &                                           &                                    &                                                &                                           & $\blacksquare$ \\
\multicolumn{1}{l|}{Developer}    & requester\_succ\_rate         & Percentage of the submitter’s PRs got merged previously.        & $\square$                                  & $\blacksquare$                                 &                                           & $\square$                          &                                                &                                           & $\blacksquare$ \\
\multicolumn{1}{l|}{Developer}    & followers                     & Followers to the developer at creation                                                         & $\square$                                  &                                                & $\square$                                 & $\square$                          &                                                &                                           &                \\
\multicolumn{1}{l|}{Developer}    & collaborator\_status          & The user's collaborator status within the project                                              &                                            & $\square$                                      & $\blacksquare$                            & $\square$                          &                                                &                                           & $\square$      \\
\multicolumn{1}{l|}{Developer}    & experience                    & Developers‘ working experience with the project                                                &                                            &                                                &                                           &                                    &                                                & $\blacksquare$                            & $\square$      \\ \hline
\multicolumn{1}{l|}{Other}        & Friday effect                 & True if the pull request arrives Friday                                                        &                                            &                                                &                                           & $\square$                          &                                                &                                           &               
\end{tabular}

\end{adjustbox}

\end{table*}
 
\section{Methodology}
\label{methods}
    To leverage the advantages of crowdsourcing, we perform a five-step process from exploring prior studies to running the replication studies using crowdsourcing and then data mining.  
    
    \noindent {\bf Step 1:} {\em Related work exploration on GitHub pull requests studies; extract data, insights, features and results from the existing work. Quantitative features should also be extracted from existing work (to answer RQ3).}
    
    \noindent {\bf Step 2:} {\em Map insights from qualitative work into questions that could be answered by crowd workers in micro-tasks. Map existing quantitative features into questions with known answers, which are ``gold" questions used for quality control. }
    
    \noindent {\bf Step 3:} {\em Collect more artifacts using similar sampling processes to the primary study. Apply the mapped questions from Step~2 to the additional data.}
    
    \noindent {\bf Step 4:} {\em Using the original data as ``gold" queries for quality control in crowdsourcing, run the crowdsourced study. }
    
    \noindent {\bf Step 5:} {\em Extract and analyze features defined in Step~2 from the crowd answers. Compare those with the findings from Step~1  to discover new insights.}

    Next, we apply these methods to the primary TDH study~\cite{tsay2014let}.

\begin{table*}[!b]
\centering
\caption{Questions for each pull request in secondary study}
\label{studyquestions}
\begin{adjustbox}{max width=\linewidth}
\begin{tabular}{|l|l|l|l|}
\hline
\textbf{Concepts}                                                                                                                                                                                       & \textbf{Questions}                                                                                                    & \textbf{Response} & \textbf{Identifier}        \\ \hline
                                                                                                                                                                                                        & Support showed                                                                                                        & Yes/No            & Q1\_support                \\
                                                                                                                                                                                                        & Support from core members                                                                                             & Yes/No            & Q1\_spt\_core              \\
\multirow{-3}{*}{\begin{tabular}[c]{@{}l@{}}Q1: Is there a comment showing support for this \\ pull request, and from which party?\end{tabular}}                                                        & Support from other developers                                                                                         & Yes/No            & Q1\_spt\_other             \\ \hline
                                                                                                                                                                                                        & Alternate solutions proposed                                                                                          & Yes/No            & Q2\_alternate\_solution    \\
                                                                                                                                                                                                        & Alternate solution proposed by core members                                                                           & Yes/No            & Q2\_alt\_soln\_core        \\
\multirow{-3}{*}{\begin{tabular}[c]{@{}l@{}}Q2: Is there a comment proposing alternate \\ solutions, and from which party?\end{tabular}}                                                                & Alternate solution proposed by other developers                                                                       & Yes/No            & Q2\_alt\_soln\_other       \\ \hline
                                                                                                                                                                                                        & Disapproval for the solution proposed                                                                                 & Yes/No            & Q3\_dis\_solution          \\
                                                                                                                                                                                                        & Disapproval due to bug                                                                                                & Yes/No            & Q3\_dis\_soln\_bug         \\
                                                                                                                                                                                                        & Disapproval because code could be improved                                                                            & Yes/No            & Q3\_dis\_soln\_improve     \\
\multirow{-4}{*}{\begin{tabular}[c]{@{}l@{}}Q3: Did anyone disapprove the proposed solution \\ in this pull request, and for what reason?\end{tabular}}                                                 & Disapproval due to consistency issues                                                                                 & Yes/No            & Q3\_dis\_soln\_consistency \\ \hline
                                                                                                                                                                                                        & Disapproval for the problem being solved                                                                              & Yes/No            & Q4\_dis\_probelm           \\
                                                                                                                                                                                                        & Disapproval due to no value for solving this problem                                                                  & Yes/No            & Q4\_dis\_prob\_no\_value   \\
\multirow{-3}{*}{\begin{tabular}[c]{@{}l@{}}Q4: Did anyone disapprove the problems being \\ solved? E.g., question the value or appropriateness \\ of this pull request for its repository.\end{tabular}} & \begin{tabular}[c]{@{}l@{}}Disapproval because the problem being solved \\ does not fit the project well\end{tabular} & Yes/No            & Q4\_dis\_prob\_no\_fit     \\ \hline
Q5: Does this pull request get merged/accepted?                                                                                                                                                         & Pull request got merged into the project                                                                              & Yes/No            & Q5\_merged                 \\ \hline
\end{tabular}
\end{adjustbox}
\end{table*} 
    
    \subsection{Step 1: Literature Overview and Data Extraction} \label{literature review}

     We first identified TDH as our primary study after finding its data source is public available and some of its insights about GitHub pull requests could be mapped into quantitative questions for the crowd to answer. 
    Next, we explored prior work related to GitHub pull requests in the literature.

    For the literature exploration, we searched for keywords `pull', `request' and `GitHub' on Google Scholar from 2008 to 2016 and also obtained a dataset from  16 top software engineering conferences, 1992 to 2016~\cite{mathew17}, filtering out the work unrelated to GitHub pull requests. 
    Table~\ref{tab:lit_rev} lists the remaining research papers that have studied pull requests in GitHub using either qualitative or quantitative methods. 
    Here, we distinguish qualitative and quantitative methods by whether or not there is human involvement during data collection process. Qualitative studies have human involvement and include interviews, controlled human experiments, and surveys. 
    We observe that all previous studies on pull request in GitHub, except for one, use either qualitative or quantitative methods.
    The remaining study combines both with a very time consuming manual analysis for the qualitative part~\cite{kalliamvakou2014promises}, which starts from the very beginning with no previous knowledge. This is quite different from ours; we leverage prior work and apply crowdsourcing directly on the results extracted from primary qualitative studies. 
    
     Table~\ref{all_features} summarizes the features found to be relevant in determining pull request acceptance. This includes all quantitative  papers from Table~\ref{tab:lit_rev} that  use features to predict the outcomes of pull requests, and the features explored in at least one of those papers.
      In  Table~\ref{all_features}:
    \begin{itemize}
    \item
    White boxes $\square$ denote that a paper examined that feature;
    \item
    Black boxes $\blacksquare$ denote when  that paper concluded that feature was important;
     \end{itemize}
    The last column in Table~\ref{all_features} shows what lessons we took from these prior studies for the data mining analysis (RQ3). If any other column marked a feature as important, then we added it into the set of features we examined. Such features are denoted with a white box $\square$  in the last column.
    If, in RQ3, we  determine the feature is informative for pull request acceptance, it is marked with a black box $\blacksquare$.

     \subsection{Step 2: Map Insights into Questions and Features} \label{Task Template}

    The tasks performed by the crowd were designed to collect quantitative information about the pull requests, which could be checked against a ground truth extracted programmatically (e.g., was the pull request accepted?), and also collect information related to the pull request discussion, which cannot be easily extracted programmatically, described next.
    
    The primary study~\cite{tsay2014let} concluded, among other things, that:
    \begin{quote}
		\noindent \textit{Methods to affect the decision making process for pull requests are mainly by \underline{offering support}~(Q1) from either external developers or core members.}\\
		
			\noindent\textit{Issues raised around code contributions are mostly \underline{disapproval for the problems being solved}~(Q4), \underline{disapproval for the solutions}~(Q3) and \underline{suggestion for alternate solutions}~(Q2).}
	\end{quote}
	These are the insights we use to derive quantitative questions for the crowd, which are mapped to the question in Table~\ref{studyquestions} (including Q5 regarding pull request acceptance). 
	
		In order to use crowdsourcing to do a case study for pull requests, our tasks contained questions related to the four concepts underlined above and  shown explicitly in Table~\ref{studyquestions} in the \emph{Concepts} column. This is followed by the \emph{Questions} related to each concept. For example, in Q2, the worker would answer \emph{Yes} or \emph{No} depending on whether alternate solutions were proposed at all (\emph{Q2\_alternate\_solution}), were proposed by core members (\emph{Q2\_alt\_soln\_core}), or were propose by other developers (\emph{Q2\_alt\_soln\_other}). 
	These four concepts reference  important  findings from TDH's work and were selected because they could be easily converted into micro questions for crowd workers to answer, though we note that not all the TDH findings were converted into questions for the crowd.  
	The full version of our questions are available on-line (http://tiny.cc/mt\_questions).

 Per Step~2 in our methods (Section~\ref{methods}), we use quantitative questions over the original pull requests from the TDH study as gold standard \emph{tasks}. 
    After extracting answers from the TDH results, we compare the crowd's performance on those pull requests to ensure the crowd is qualified to perform the tasks. 
    
     To further ensure response quality in a crowdsourced environment, for all pull requests,  we also added three  preliminary qualification questions that require crowd workers to identify the submitter, core members and external developers for each pull request; these are gold standard \emph{questions}. 
    These extra questions let a crowd worker grow familiar with analyzing pull request discussions, and let us reject
	 answers from unqualified crowd workers since we could programmatically  extract the ground truth from the pull request for comparison. Details on our quality control used during the study are in Section~\ref{crowddata}.

     \subsection{Step 3: Data Collection and Expansion}
    \label{pr selection}
    
      To make sure the pull requests are statistically similar to those of TDH's work~\cite{tsay2014let}, we applied similar selection rules on 612,207 pull requests that were opened  as in new in January 2016 from GHTorrent~\cite{Gousi13GHTorrent}, which is a scalable, searchable, offline mirror of data offered through the GitHub Application Programmer Interface (API). The selection criteria are stated as follows. The main difference between our selection and TDH's selection is the time requirements for when the pull requests are created or last updated.
    \begin{enumerate}
      \item Pull requests should be closed. 
      \item Pull requests should have comments. 
      \item Pull request comment number should be above eight.
      \item Exclude pull requests whose repository is a fork to avoid counting the same contribution multiple times.
      \item Exclude pull requests whose last update is not later than January, 2016, so that we can make sure the project is still active.
      \item Retain only pull requests with at least three participants and where the repository has at least ten forks and ten stars.
    \end{enumerate}
    There are 565 pull requests left after applying the selection criteria stated above. From these pull requests, we sampled \newPR such that half were ultimately merged and the other half were rejected.
    
     The \newPR additional pull requests were
    published on MTurk for analyzing in two rounds, together with the 20 carefully studied pull requests from TDH~\cite{tsay2014let} inserted for each round as ``gold" standard tasks. The 1st round has 100~(80 new + 20 TDH) pull requests in total, while 2nd round has 110~(90 new + 20 TDH) pull requests in total. So \totalPR total pull request were published on for the crowd to complete. 

    \subsection{Step 4: Crowdsourcing Study} \label{quality control in cs}
    \label{crowddata}
    \label{Quality Control}
    Mechanical Turk is a crowdsourcing platform that connects workers (i.e., the people performing the tasks) with requesters (i.e., the people creating the tasks)~\cite{mturk}.  MTurk has been used extensively in software engineering applications~\cite{mao15} and for the evaluation of software engineering research (e.g., \cite{Stolee:2010:EUC:1852786.1852832, stoleeesem2013, Fry:2012:HSP:2338965.2336775}). 
    It manages recruitment and payment of participants and hosts the tasks posted by requesters. 
    
    \subsubsection{Tasks}
    \label{eachtaskquestions}
    The tasks performed by participants are called Human Intelligence Tasks (HITs). These represent individual activities that crowd workers perform and submit. The scope of a HIT for our study included a single pull request, the questions outlined in Table~\ref{studyquestions}, as well as the following questions about the pull request that could be checked programmatically for quality control: 
    \begin{enumerate}
      \item What is the submitter GitHub ID?
      \item From the contributor page of this repository, who are the core members of this repository?
      \item Who are the external developers among the participants of this pull request?
      \item Does this pull request get merged/accepted?~(Q5)
    \end{enumerate}
    
    \subsubsection{Participants}
    Workers for our study were required to have demonstrated high quality in prior tasks on the MTurk platform and answer simple questions about the pull request correctly. 
    Quality and experience filters were applied to screen potential participants; only workers with HIT approval rate above 90\%, and who had completed at least 100 approved HITs could participate. 
     To make sure the crowd participants are qualified to analyze the pull requests in our study, we also require them to be GitHub users.

    \subsubsection{Cost}
     \label{Cost Control}
    We want to make sure the cost is low but also provide a fair payment for the participants. According to several recent surveys on MTurk~\cite{buhrmester2011amazon, berinsky2012evaluating, paolacci2010running, ipeirotis2010demographics}, the average hourly wage is \$1.66 and MTurk workers are willing to work at \$1.40/hour. We estimated about 10 minutes needed for each HIT, and first launched our task with \$0.25 per HIT but only received 1 invalid feedback after 2 days. So we doubled our payment to \$0.50 for each HIT, which requires to analyze one single pull request.  Each round of tasks were completed in one week. Our final results show that 17 minutes are spent for each HIT on average, which means \$1.76 per hour. In total, 27 workers participated in our tasks, and 77 hours of crowd time were spent to get all the pull requests studied. 
    
    \subsubsection{Quality Control}
    A major issue in crowdsourcing
    is how to reduce the noise inherent
    in data collection. This section
    describes the three approaches 
    we used to increase data quality:
    \begin{itemize}
        \item  Qualification questions; 
        \item Redundant question formats~\cite{stolee2015exploring}; and 
        \item ``Gold" standard tasks~\cite{alonso2011design, sarasua2012crowdmap}.
    \end{itemize}
    
    \noindent \paragraph{Qualification Questions}
    A domain-specific screening process was applied as participants were required to answer preliminary qualification questions related to identifying the pull request key players and pull request acceptance {\em on every pull request analyzed}. These are questions for which we can systematically extract values from the pull requests (i.e., those in Section~\ref{eachtaskquestions}); if these objective questions are answered incorrectly, the task was rejected and made available to other crowd workers.  
    
    This is unlike the {\em qualification test} which is available on the MTurk platform to screen participants once for eligibility to participate in any of our tasks. A qualification test is administered once. Once participants pass, they can perform our HITs. Our {\em qualification questions}, on the other hand, were used for every HIT we published. This ensured quality responses for each and every HIT.
    
    \paragraph{Redundant Question Formats}
    For each question in the task related to the pull request comment discussion, we require workers to answer a yes/no question and then copy the comments supporting their answers from the pull request into the text area under each question, as suggested in prior work~\cite{stolee2015exploring}. Take question 1 for example (\textit{Is there a comment proposing alternate solutions?}): if they choose \textit{"Yes, from core members"}, then they need to copy the comments within the pull requests to the text area we provided.

    \noindent \paragraph{``Gold" Standard Tasks}
    This study was run in two phases. In each phase, the original 20 pull requests were added to the group as ``gold" standard tasks. 
    The tasks were randomly assigned to crowd workers. 
    For those crowd workers who got one of the 20 previously studied pull requests, we checked their answers against the ground truth~\cite{tsay2014let}; inaccurate responses were rejected and those workers were blocked. This acted as a random quality control mechanism.  
    
    We used multiple quality control mechanisms in keeping with prior work~\cite{pengSANER2019}, and the result quality was satisfactory. However, 
    evaluating the effectiveness of each quality control mechanism we employed is left for future work. 
    
      \subsection{Step 5: Data Mining Analysis} \label{feature_selection}
    
    In this study,  we have two groups of features: (1)~all the quantitative features found important in previous works (see Table~\ref{all_features}), and (2)~the qualitative features extracted from the results of studying pull requests in detail by the qualified crowd (see Section~\ref{crowddata}). For each group of features, we run the CFS feature selector~\cite{hall1999correlation} to reduce the features to use for our decision tree classifier.    
  To collect the quantitative features, we used the GitHub API
  to extract the features marked in the   \emph{ours} column of Table~\ref{all_features}.
  
    CFS evaluates and ranks feature subsets.
     One reason to use CFS over, say, correlation, is that CFS
   returns {\em sets} of useful features while simpler
   feature selectors do not  understand the interaction between
   features.
   CFS also assumes that a ``good'' set of features   contains features that are highly connected with the target class, but weakly connected to each other. 
    To implement this heuristic,
    each feature subset is scored as follows according to Hall, et al.~\cite{hall1999correlation}:
    \[
    \mathit{merit}s = \frac{k\overbar{r_{\mathit{cf}}}}{ \sqrt{k+k(k-1)\overbar{r_{\mathit{ff}}}}}
    \]
    The $\mathit{merit}s$ value is some subset $s$ containing $k$ features;
    $r_{\mathit{cf}}$ is a score describing the connection of that feature
    set to the class;
    and $r_{\mathit{ff}}$ is the mean score of the feature to feature
    connection between the items in $s$.
    Note that for this fraction to be maximal, $r_{\mathit{cf}}$ must be large
    and $r_{\mathit{ff}}$ must be small, which means features have to correlate
    more to the class than each other.
    
    This equation is used to guide a  best-first search with a horizon of five  to select most informative set of features. 
    Such a search proceeds as  follows. The initial frontier is all sets containing one different feature. The frontier of size $n$, which initialized with $1$, is sorted according to {\em merit} and the best item is grown to all sets of size {\em n+1} containing the best item from the last frontier. The search
    stops when no improvement have been seen in last five frontiers in $merit$. Return the best subset seen so far when stop.

      Our experiments assessed
three groups of features:
\begin{enumerate}
\item
    After CFS feature selector, the selected {\em quantitative features} are 
    \textit{commits\_on\_files\_touched, 
    requester\_succ\_rate, 
    prev\_pullreqs}, 
    which are quite intuitive. (The {\em commits\_on\_files\_touched} feature indicates the popularity of the modified files is highly relating to the final acceptance of the pull request while {\em requester\_succ\_rate} and {\em prev\_pullreqs} show the requester's history interaction with the project also play an important role for the final acceptance.)
    \item
    
    Using the same CFS feature selector, the selected {\em crowdsourced features} are
    \textit{Q3\_dis\_solution, \newline
    Q4\_dis\_prob\_no\_value}. These two selected feature show that the final acceptance has a strong relation to the opposing voice in the pull requests discussion, especially for comments saying this pull request has no value for the project.
    \item 
    Combining all {\em quantitative features} and  {\em crowdsourced features} into {\em combined
    features} and feed that into the same CFS feature selector, the following features were selected:
    \textit{ Q3\_dis\_solution, 
    commits\_on\_files\_touched, 
    requester\_succ\_rate,} and 
    \textit{prev\_pullreqs}, which all appeared using the CFS selector on the quantitative and qualitative features independently.  
\end{enumerate} 

For each of these three sets of features, we  ran a 10x5 cross validation for supervised learning with the three different groups of features. 
These generate three models that predicted if a pull request would get merged/accepted or not. 
A decision tree learner was used as our supervised learning algorithm.  This was selected
after our initial studies with several other learners that proved to
be less effective in this domain (Naive Bayes and SVM).

Using the MTurk micro-task crowdsourcing platform, we collect data for 1) the original 20 pull requests from the primary study (twice), and 2) \newPR additional, independent pull requests. 
Of the \uniquePR unique pull requests, 176  pull requests were evaluated by the crowd with high quality. Of those, 156 are new pull requests (dropping 14 from the sample of 170) and 34 of them are from the original TDH study (covering each of the 20 original pull requests at least once). 
    The unqualified responses were a result of the redundant question format quality control approach, but an operational error led us to approve the tasks despite the poor comment quality, leaving us with a smaller data set for further analysis.

     The crowd data includes qualitative information about the pull request discussion, such as whether there is a comment showing support, proposing an alternate solution, disapproving of the solution, and disapproving of the problem being solved.
    We refer to the data collected via crowd as the {\em qualitative} pull request features.

    \section{Results}
    The crowd data is substantially larger than the original data in terms of pull requests analyzed. 
    The benefits of the larger sample size is two-fold. 
    First, by using similar selection criteria in the replication study compared to the primary study, we are able to check the 
    stability and external validity of the findings in the primary study using a much larger sample (RQ2). 
    Second, in terms for informativeness, we can extract features from crowd's answers, 
    which is qualitative, and build models to predict pull request acceptance results. 
    This allows us to compare the performance of models built with (a)~some features identified as important in the primary study and (b)~the features from related, quantitative works (RQ3).
    
    \subsection{RQ1: Can the crowd reproduce prior results with high quality?} \label{RQ_1}

\begin{table}[!t]
\centering
\caption{Quality for Crowdsourcing Results from Amazon Mechanical Turk (RQ1).}
\label{mt_quality}
\begin{tabular}{|l|l|l|l|}
\hline
\textbf{Questions} & \textbf{Precision} & \textbf{Recall} & \textbf{F1-Score} \\ \hline
\textbf{Q1}        & 0.769               & 0.769           & 0.770             \\ \hline
\textbf{Q2}        & 0.818               & 0.750           & 0.783             \\ \hline
\textbf{Q3}        & 0.727               & 0.667           & 0.696             \\ \hline
\textbf{Q4}        & 0.778               & 0.700           & 0.737             \\ \hline
\textbf{Q5}        & 0.833               & 0.714           & 0.770             \\ \hline
\textbf{Total}     & \textbf{0.801}      & \textbf{0.742}  & \textbf{0.770}    \\ \hline
\end{tabular}
\end{table}

   RQ1
   checks if our mapping in Section~\ref{Task Template} correctly captures the essence of the TDH study.
   Here, we used the results from the original 20 pull requests from TDH. 
   
    Table \ref{mt_quality} shows the   $precision$, $recall$ and $F_1$ scores of the crowd working
    on the original tasks, using the prior TDH study as the ground truth. The {\em Questions} map to Table~\ref{studyquestions}. As    seen in Table~\ref{mt_quality}, the precision and recall of the crowd on the 20 original tasks is 80\% and 74\% respectively (so  $F_1 \approx 77\%$). 
    Based on prior
    work with data mining from software engineering data~\cite{bird2015art, Fu2016TuningFS}, we find that these values
    represent a close correspondence between the TDH results and those from the crowd, hence indicating that the crowd can indeed perform the tasks with high quality.
    
     We further manually examined the cases where the crowd disagreed with  TDH, finding that sometimes TDH appears correct, and sometimes the crowd appears correct. For example, TDH classifies the 17th pull request they studied as no support, while the crowd found the comment from the user {\tt drohthlis} saying {\em ``This is great news!"},  which is an apparent indicator for the supporting this pull request after our examination. 
    Another two cases are the 16th and 20th pull requests they studied. Crowd workers found  clear suggestions for alternative solutions (i.e., {\em ``What might be better is to ...''}, {\em ``No, I think you can just push -f after squashing.''}), which TDH does not find. 
    While we did throw away some data due to quality, largely the crowd was able to replicate, and extend, some of the original TDH study.

    As to the issues
of speed and cost, 
    TDH report that they required about 47 hours to collect interview
     data on 47 users within which, they investigated the practices about pull requests. TDH does not report the subsequent analysis time. By way of comparison, 
   we spent \$200, to buy 77  hours of crowd time. In that time,  \uniquePR (156 new ones and 34 from the primary study, which covering all 20 original pull requests) out of \totalPR published pull requests were validly analyzed. 
   
   In summary, we answer RQ1 in the affirmative. 
    
     \subsection{RQ2: Are the primary study's results stable? } \label{RQ_2}

    \begin{figure}
    \centering
    \includegraphics[width=\columnwidth]{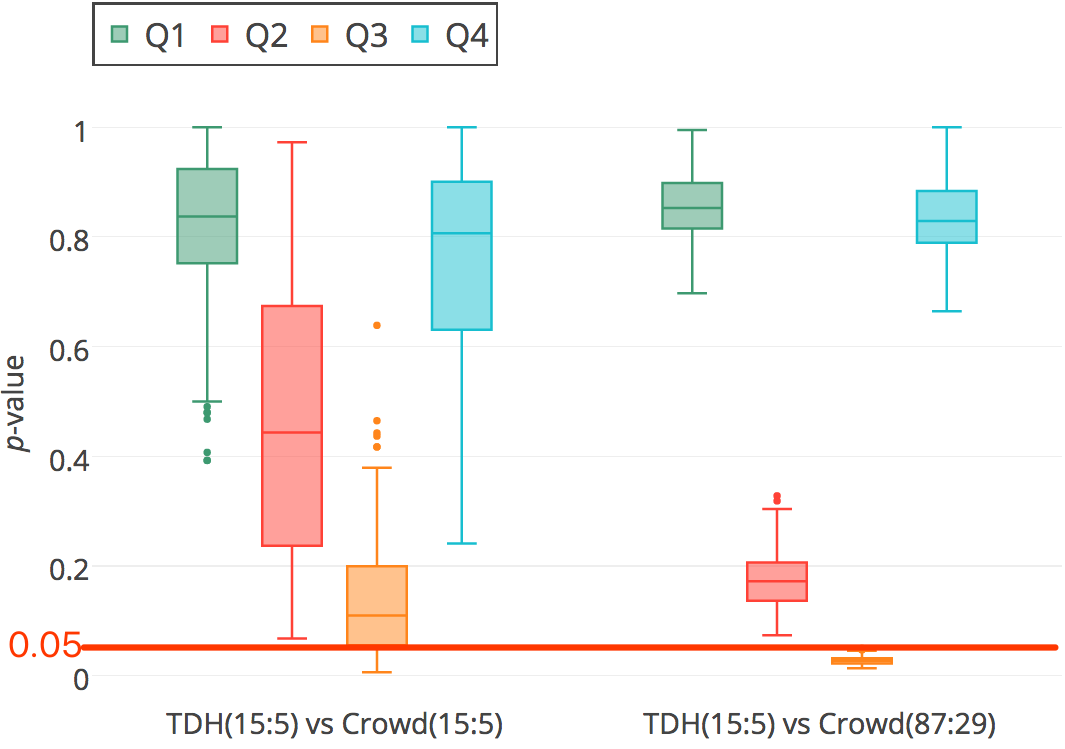}
    \caption{Stability Checking: \textit{p}-values for comparing unscaled and scaled crowd answers with answers from TDH}
    \label{fig:stability_box}
    \end{figure}
    
     One motivation for this work was
     checking if crowdsourcing can scale and confirm the  external validity of
     qualitative conclusions. 
     This issue is of particular concern for crowdsourcing  studies due the subjective nature
     of the opinions from the crowd. If those opinions {\em increased} the variance of the collected
     data, then the more data we collect, the {\em less} reliable the conclusions.
     
     To test for this concern, we compare the pull requests studied by crowd (excluding the 20 gold tasks) with the 20 pull request studied by TDH. Among the original 20, 15 were merged and 5 were rejected. Given the 156 new pull requests, we first randomly select 15 merged and 5 rejected pull requests  100 times, so that we can compare the these 2 independent samples at the same scale and with the same distribution of pull request acceptance outcomes. Then we run another 100 iteration for randomly selecting 87 merged and 29 rejected pull requests studied by crowd, which still has the same distribution but at nearly six times larger scale. \textit{p}-values are collected for each sample comparison in the 2 runs. We analyzed each of the four concepts from Table~\ref{studyquestions} separately (Q1-Q4).

     Figure~\ref{fig:stability_box} shows the results of comparing pull requests from TDH and an independent sample with 2 different scales. As shown, Questions 1, 2, 4 are quite stable for both scales (i.e., p-values greater than 0.05 indicate no statistically significant difference with $\alpha = 0.05$). Moreover, Question 1 and 4 are becoming more stable when scale becomes larger, while Question 2 becomes less stable at a larger scale. For Question~3, all of the p-values are lower than 0.05 at the large scale, though the median of its p-value is higher than 0.05 at the same scale as TDH. 
     This may indicate that TDH did not cover enough pull requests to achieve a representative sample for the finding, which is mapped into Question 3 about disapproving comments. 
     This may also indicate that TDH treated disapproving comments in GitHub pull requests differently than the crowd. 
     As we have shown when answering RQ1, we have found several cases where the crowd and TDH disagree and the crowd is correct, though the crowd did have some wrong answers while TDH are correct.
     Note that the results are not to fault TDH, but serves as the evidence why we need to replicate and scale empirical studies.

     Accordingly, we answer RQ2 in the affirmative. The results for Q1, Q2, and Q4 do not differ significantly between TDH and independent samples of the same size or of a larger size. The exception is Q3, for which the results differ significantly when scaling to a larger data set.

                     \begin{figure}
    \centering
    \includegraphics[width=\columnwidth]{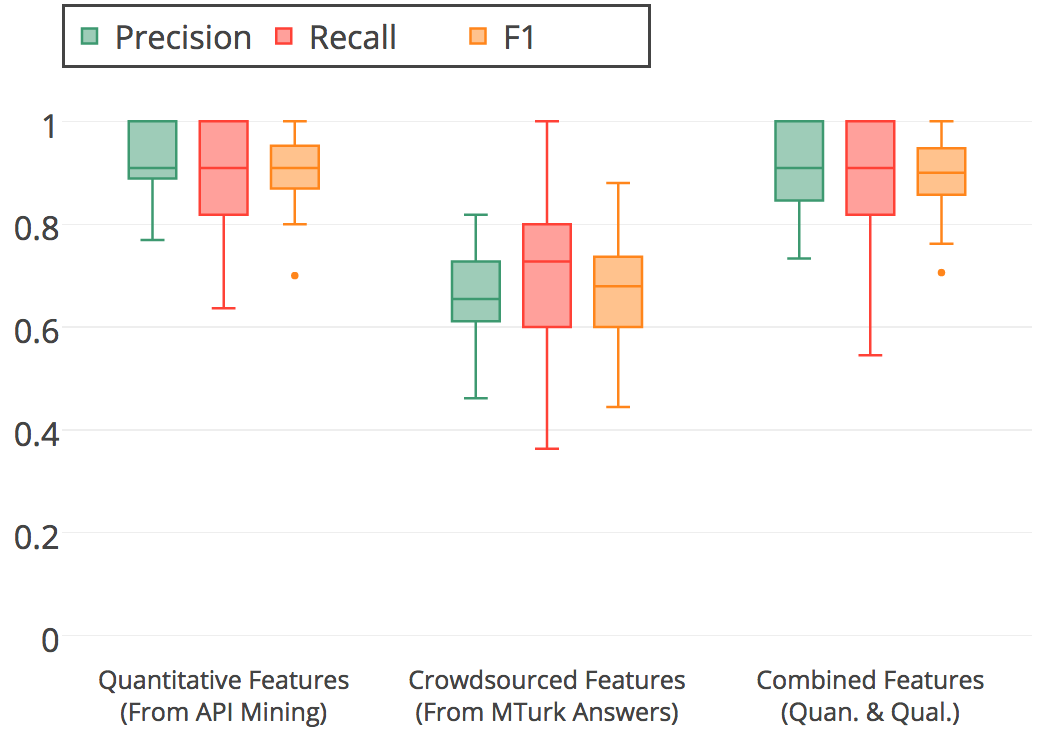}
    \caption{Performance Comparison for Using different feature set to predict whether a pull requests will be accepted.}
    \label{fig:performance}
    \end{figure}
    
      \subsection{RQ3: How well can the qualitative and quantitative features predict PR acceptance?} \label{RQ_3}

     The results are shown in Figure \ref{fig:performance}, expressed
    in terms of precision, recall, and the $F_1$
    score; i.e.
    the harmonic mean
    of precision and 
    recall, for each of three feature sets: quantitative, crowdsourced, and a combination. 
    Note that  the
    performances of the predictor using crowdsourced features 
    are not as high or as stable as the one built with quantitative features. 
    We can see that:
    \begin{itemize}
        \item The selected quantitative features achieved $F_1$ score at 90\% with a range of 20\%;
        \item The selected crowdsourced features achieved lower $F_1$ score at 68\% with a larger range.
        \item The combined selected features did better than just using qualitative; but performed no better than just using only the quantitative features.
    \end{itemize}
    
     At first glance, the models learned from crowdsourced features
   performed worse than using  quantitative features extracted from numerous prior
   data mining studies, but this warrants further discussion.  
   While  Figure~\ref{fig:performance} shows the crowdsourced features generated from TDH results perform well on predicting the final acceptance with $F_1$ at 68\%, it might be that considering more of the features from the original study could lead to even better results. 
   For our study, we performed an independent partial replication and selected features that could be easily mapped to micro-questions, as stated in Section~\ref{Task Template}. 
   That said, the instability observed for Q3 as the sample size increases (Figure~\ref{fig:stability_box}) may be present for other features from the original study; further study is needed to expand the number of questions asked of the crowd to tease out this phenomenon. 
   
   In summary, the quantitative features extracted from dozens of recent papers on pull requests outperform the selected qualitative features we studied.

\section{Threats to Validity and Limitations}

As with any empirical study, biases and limitations can affect the final
    results. Therefore, any conclusions made from this work must
    be considered with the following issues in mind.

 \textbf{Sampling bias:} This threatens the external validty of any classification experiment.
         For example,
        the pull requests used here are selected using the rules described in \ref{pr selection}. Only \newPR additional  highly discussed pull requests from active projects are sampled and analyzed, so our results may not reflect the patterns for all the pull requests. That said, we note that one reason to endorse crowdsourcing is that its sample size can be much larger than using just qualitative methods.  For example, TDH reported results from just 20 pull requests.
        
  \textbf{Learner bias:} For building the acceptance predictors in this
        study, we elected to use a decision tree classifier. We chose   decision trees
        because it suits for small data samples and its results were comparable to 
        the more complicated algorithms like Random Forest and SVM. Classification is a large and active field
        and any single study can only use a small subset of the known
        classification algorithms. Future work should repeat this study using other learners.
        
   \textbf{Evaluation bias:} This paper uses \textit{precision}, \textit{recall} and  $F_1$ \textit{score} measures   of predictor's performance. Other performance measures used in software engineering
        include accuracy and entropy.  Future work should repeat this study using different evaluation biases.

     \textbf{Order bias:} For the performance evaluation part, the order that the data trained  and predicted affects the results. To mitigate this order bias, 
      we run the 5-bin cross validation 10 times randomly changing the order of the pull requests each time.

      \section{Implications}   \label{implications}

    This paper presents one example in which crowdsourcing and data mining were used for partial replication and scaling of an empirical study. 
    In our work, we use the experts to identify a set of interesting questions, and the crowd to answer those questions regarding a large data set. Then, data mining was used to compare the qualitative responses to quantitative information about the same data set. 
    It is possible this workflow is indicative of a more general framework for scaling and replicating qualitative studies using the approach we used. 

    In the original qualitative TDH study, the authors found that 1)~\textit{supports}, 2)~\textit{alternate solutions}, 3)~\textit{disapproval for the proposed solutions}, and 4)~\textit{disapproval for the problems being solved} were important factors that guard pull requests' acceptance. In this scaled, crowdsourced replication, we found that factors 1, 2 and 4 still hold, but 3 was unstable. As shown in section \ref{RQ_2}, we cannot tell whether the crowd are correct or TDH are correct, because we have observed situations where one is correct while the other is not for both sides. However, the implication here is that the combination of empirical methods allows us to pinpoint more precisely results that are steadfast against tests of external validity and the results that need further investigation.
    
    In the end, the replication via quantitative study would have been impossible without the primary qualitative work, and we should make best use of  time-consuming qualitative work, instead of stopping after we get results from qualitative results (and vice versa).
    We find qualitative studies can inspire quantitative studies by carefully mapping out areas of concern.  Primary  qualitative  study  can also  provide  the  data  needed  to direct replications via quantitative crowdsourcing studies. We also find a single  primary  qualitative  study  can  direct  the  work  of many quantitative studies, and our work is just one example of the a partial replication study after TDH's qualitative work.

\section{Related Work}
  \label{sec_related_work}
    There are many ways to categorize empirical studies in software engineering. 
    Sjoberg, et al.~\cite{sjoberg2007future} identify two general groups, primary research and secondary research.
    The most common primary research 
    usually involves the collection and analysis of original
    data, utilizing methods such as experimentation,
    surveys, case studies, and action research. 
    Secondary studies involve synthesis and aggregation of primary studies. 
    According to Cohen~\cite{cohen1989developing}, secondary studies can identify
    crucial areas and questions that have not been addressed
    adequately with previous empirical research.

    Our paper falls into primary research that 
    uses data from previously published 
    studies. 
    Using independent, partial replication, we check the external validity of some of the TDH results. 
    That is, we explore  whether claims for generality  are
    justified, and  whether our study yields the same results if we partially replicate it~\cite{easterbrook2008selecting}.
    Our data collection applies the coding method, which is commonly used to extract values for
    quantitative variables from qualitative data in order to perform some
    type of quantitative or statistical analysis, with crowdsourcing to manage participant recruitment. 
    Most of our data analysis falls into the category of replication for theory confirmation as we independently replicate parts of the results from TDH. However, in the data mining study, we introduce additional factors not studied in the original TDH study.

In software engineering, crowdsourcing is being used for many tasks traditionally performed by developers~\cite{mao2015survey} including
        program synthesis~\cite{Cochran:2015:PBP:2676726.2676973},  
        program verification~\cite{Schiller:2012:RBW:2398857.2384624}, 
        LTL specification creation~\cite{pengSANER2019},
        and  
        testing~\cite{6569745, Nebeling:2013:CGT:2494603.2480303}. According to Mao, et al.~\cite{mao2015survey}, our work falls into the evaluation of SE research using crowdsourcing as the crowd is not performing software engineering tasks, but rather is used to evaluate software artifacts. 
        


There are many reasons to use crowdsourcing in software engineering, and this space has been studied extensively~\cite{mao2015survey}.
Two important issues with crowdsourcing are        {\em quality control} and {\em cost control}.
While the crowd is capable of producing quality results, systems need to be in place to monitor result quality. Cost has a dependent relationship on the quality control measures; a lack of quality control can lead to results that must be thrown out and repeated, hence increasing the cost.

{\em Quality control} in crowdsourcing has been studied extensively within and beyond software engineering applications (e.g.,~\cite{alonso2011design, sarasua2012crowdmap, stolee2015exploring, krosnick1991response, shiel2013conflict, Kim:2012:FOR:2442576.2442591}). 
        One way to address quality control is to use a golden set of questions~\cite{alonso2011design, sarasua2012crowdmap} for which the answer is known.  Workers that perform poorly on the golden set are eliminated, which is also one of the strategies we take in this paper. 
        Alternatively, tasks can be assigned to multiple workers and their results aggregated (e.g., in the TURKIT system~\cite{5295247}, one task
        is performed iteratively, and each worker is asked to improve on the answer of the former, or in AutoMan~\cite{Barowy:2012:API:2398857.2384663}, each task is performed by crowd members until statistical consensus is reached). 
        The tradeoffs are in cost and volatility. The gold questions can be checked automatically, yielding low cost, but there may be high volatility in the results from a single individual. When using multiple people and aggregating the results, volatility is lower as the crowd is likely to converge~\cite{minku15}, but the cost will be higher due to the required replication per question. 
        Other quality control techniques include redundant question formats~\cite{stolee2015exploring}, notifications to workers about the value of their work~\cite{krosnick1991response}, answer conflict detectors~\cite{shiel2013conflict}, and random click detection~\cite{Kim:2012:FOR:2442576.2442591}.
        
        {\em Cost control} is important as commercial crowdsourcing platforms are not free. Economic incentives for crowd workers strongly effect  crowd response quality~\cite{mason2010financial, goel2014mechanism, mao2013volunteering, kittur2008crowdsourcing, mason2011use,Mao:2013:PCS:2486788.2486963, yin2014monetary, wang2013quality}. To keep the quality high, payments need to be high enough to entice participation~\cite{vinayak16} but low enough to keep study costs reasonable.

\section{Conclusion}
Crowdsourcing  and  data  mining  can  be  used  to effectively reduce the effort associated with the partial replication and  enhancement  of  qualitative  studies.

For example,   in this paper:
\begin{itemize}
\item
We presented an empirical study on scaling and extending a primary  study on pull request acceptance. Using the insights learned from the primary study, we designed a scaled  study with different empirical methodology. 
\item
To scale and extend the results from the original study, we used crowdsourcing. 
\item
Then, we applied data mining to determine if the qualitative features fared better or worse than quantitative features over the same artifacts. We show that, with results and data from TDH's primary study, it is possible to  map some of their insights into micro-questions for crowd workers and expand the data they studied to a larger scale. 
\item
Further, in that second study, we could build better predictors that seen before.
\end{itemize}                             
                We conjecture that this case study is representative of an underlying methodology for scaling and extending primary qualitative studies that require expert opinions.
                The long-term goal of this work is to encourage more  studies that replicate parts of primary
  qualitative studies and use the gained insights to design subsequent  studies. It is through independent  replication with different methodologies that we can move closer to building software engineering theories. For that task, collaborators would be welcomed.

\section*{Acknowledgements}
 The work is partially funded by NSF  awards \#1506586, \#1302169, and \#1645136.
    
\balance


\end{document}